\documentclass[10pt]{bmc_article}
\usepackage{cite}
\usepackage{ifthen}
\usepackage{latexsym}
\usepackage{amsmath}
\usepackage{multirow}
\usepackage{multicol}
\usepackage{color}
\usepackage{graphicx}
\usepackage{url}
\newboolean{publ}

\newenvironment{bmcformat}{\fussy\setboolean{publ}{true}}{\fussy}

\begin{document}
\begin{bmcformat}

\title{Mining Images in Biomedical Publications: Detection and Analysis of Gel Diagrams}

\author{%
Tobias Kuhn,\correspondingauthor$^1$\email{Tobias Kuhn\correspondingauthor{} -- kuhntobias@gmail.com}
Mate Levente Nagy,$^3$\email{Mate Levente Nagy --- mate.nagy@yale.edu}
ThaiBinh Luong,$^2$\email{ThaiBinh Luong -- thaibinh@gmail.com}
and
Michael Krauthammer$^{2,3}$\email{Michael Krauthammer -- michael.krauthammer@yale.edu}
}

\address{%
\iid(1)Department of Humanities, Social and Political Sciences, ETH Zurich, Switzerland.
\iid(2)Department of Pathology, Yale University School of Medicine, New Haven, CT, USA.
\iid(3)Program for Computational Biology and Bioinformatics, Yale University, New Haven, CT, USA.
}%

\maketitle

\begin{abstract}
Authors of biomedical publications use gel images to report experimental results such as protein-protein interactions or protein expressions under different conditions. Gel images offer a concise way to communicate such findings, not all of which need to be explicitly discussed in the article text. This fact together with the abundance of gel images and their shared common patterns makes them prime candidates for automated image mining and parsing. We introduce an approach for the detection of gel images, and present a workflow to analyze them. We are able to detect gel segments and panels at high accuracy, and present preliminary results for the identification of gene names in these images. While we cannot provide a complete solution at this point, we present evidence that this kind of image mining is feasible.
\end{abstract}

\ifthenelse{\boolean{publ}}{\begin{multicols}{2}}{}

\section*{Introduction}

A recent trend in the area of literature mining is the inclusion of images in the form of figures from biomedical publications \cite{yu2006bioinform,zweigenbaum2007briefbioinform,peng2008bioinform}. This development benefits from the fact that an increasing number of scientific articles are published as open access publications. This means that not just the abstracts but the complete texts including images are available for data analysis. Among other things, this enabled the development of query engines for biomedical images like the Yale Image Finder \cite{xu2008bioinform} and the BioText Search Engine \cite{hearst2007bioinform}. Below, we present our approach to detect and access gel diagrams. This is an extended version of a previous workshop paper \cite{kuhn2012smbm}.

As a preparatory evaluation to decide which image type to focus on, we built a corpus of 3\,000 figures that allows us to reliably estimate the numbers and types of images in biomedical articles. These figures were drawn randomly from the open access subset of PubMed Central and then manually annotated. They were split into subfigures when the figure consisted of several components. Figure \ref{fig:categories} shows the resulting categories and subcategories. This classification scheme is based on five basic image categories: Experimental/Microscopy, Graph, Diagram, Clinical and Picture, each divided into multiple subcategories. It shows that bar graphs (12.4\%), black-on-white gels (12.0\%), fluorescence microscopy images (9.4\%), and line graphs (8.1\%) are the most frequent subfigure types (all percentages are relative to the entire set of images).

\begin{figure*}[tb]
\begin{center}
\includegraphics[width=\textwidth]{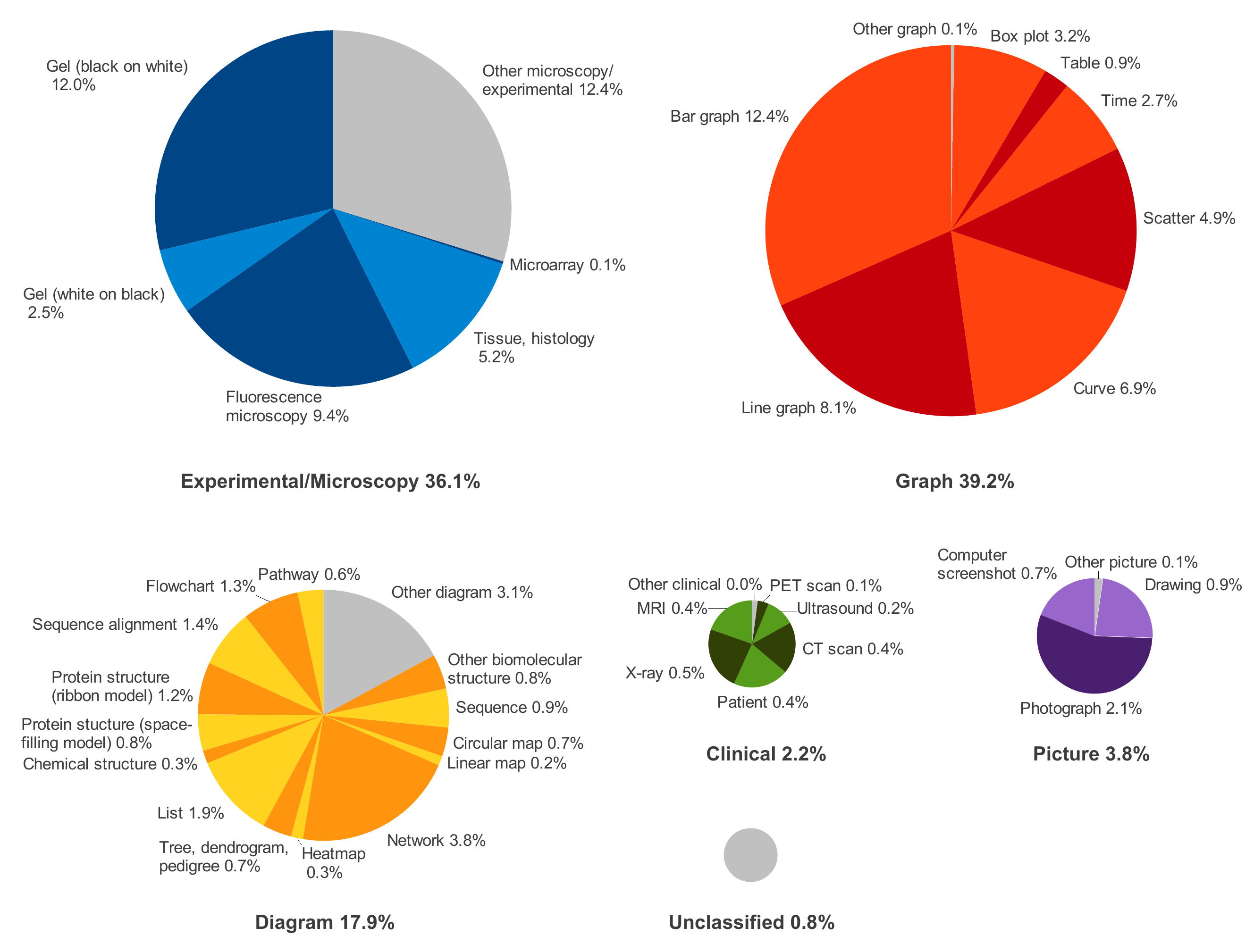}
\caption{Categorization of images from open access articles of PubMed Central.}
\label{fig:categories}
\end{center}
\end{figure*}

We targeted different kinds of graphs (i.e., diagrams with axes) in previous work \cite{kuhn2012amia}, and we decided to focus this work on the second most common type of images: gel diagrams. They are the result of gel electrophoresis, which is a common method to analyze DNA, RNA and proteins. Southern, Western and Northern blotting \cite{southern1975jmolbiol,alwine1977nas,burnette1981analbiochem} are among the most common applications of gel electrophoresis. The resulting experimental artifacts are often shown in biomedical publications in the form of gel images as evidence for the discussed findings such as protein-protein interactions or protein expressions under different conditions. Often, not all details of the results shown in these images are explicitly stated in the caption or the article text. For these reasons, it would be of high value to be able to reliably mine the relations encoded in these images.

A closer look at gel images reveals that they follow regular patterns to encode their semantic relations. Figure \ref{fig:example} shows two typical examples of gel images together with a table representation of the involved relations. The ultimate objective of our approach (for which we can only present a partial solution here) is to automatically extract at least some of these relations from the respective images, possibly in conjunction with classical text mining techniques. The first example shows a Western blot for detecting two proteins (14-3-3$\sigma$ and $\beta$-actin as a control) in four different cell lines (MDA-MB-231, NHEM, C8161.9, and LOX, the first of which is used as a control). There are two rectangular gel segments arranged in a way to form a $2\times4$ grid for the individual eight measurements combining each protein with each cell line. A gel diagram can be considered a kind of matrix with pictures of experimental artifacts as content. The tables to the right illustrate the semantic relations encoded in the gel diagrams. Each relation instance consists of a condition, a measurement and a result. The proteins are the entities being measured under the conditions of the different cell lines. The result is a certain degree of expression indicated by the darkness of the spots (or brightness in the case of white-on-black gels). The second example is a slightly more complex one. Several proteins are tested against each other in a way that involves more than two dimensions. In this case, the use of ``+'' and ``--'' labels is a frequent technique to denote the different possible combinations of a number of conditions. Apart from that, the principles are the same. In this case, however, the number of relations is much larger. Only the first eight of a total of 32 relation instances are shown in the table to the right. In such cases, the text rarely mentions all these relations in an explicit way, and the image is therefore the only accessible source.
\begin{figure*}[t]
\begin{center}
\footnotesize
\begin{tabular}{rl}
\begin{tabular}{c}
\includegraphics[scale=5]{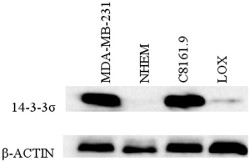}
\end{tabular}
&
\begin{tabular}{|l|l|l|}
\hline
\textbf{Condition} & \textbf{Measurement} & \textbf{Result} \\
\hline
MDA-MB-231 & 14-3-3$\sigma$ & high expression \\
\hline
NHEM & 14-3-3$\sigma$ & no expression \\
\hline
C8161.9 & 14-3-3$\sigma$ & high expression \\
\hline
LOX & 14-3-3$\sigma$ & low expression \\
\hline
MDA-MB-231 & $\beta$-actin & high expression \\
\hline
NHEM & $\beta$-actin & high expression \\
\hline
C8161.9 & $\beta$-actin & high expression \\
\hline
LOX & $\beta$-actin & high expression \\
\hline
\end{tabular}
\\
\bigskip\\
\begin{tabular}{c}
\includegraphics[scale=0.6]{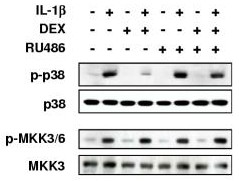}
\end{tabular}
&
\begin{tabular}{|lll|l|l|}
\hline
\multicolumn{3}{|l|}{\textbf{Condition}} & \textbf{Measurement} & \textbf{Result} \\
\hline
IL-1$\beta$ (--) & DEX (--) & RU486 (--) & p-p38 & low expression \\
\hline
IL-1$\beta$ (+) & DEX (--) & RU486 (--) & p-p38 & high expression \\
\hline
IL-1$\beta$ (--) & DEX (+) & RU486 (--) & p-p38 & no expression \\
\hline
IL-1$\beta$ (+) & DEX (+) & RU486 (--) & p-p38 & low expression \\
\hline
IL-1$\beta$ (--) & DEX (--) & RU486 (+) & p-p38 & no expression \\
\hline
IL-1$\beta$ (+) & DEX (--) & RU486 (+) & p-p38 & high expression \\
\hline
IL-1$\beta$ (--) & DEX (+) & RU486 (+) & p-p38 & low expression \\
\hline
IL-1$\beta$ (+) & DEX (+) & RU486 (+) & p-p38 & high expression \\
\hline
... & & & ... & ... \\
\hline
\end{tabular}
\\
\end{tabular}
\caption{Two examples of gel images from biomedical publications (PMID 19473536 and 15125785) with tables showing the relations that could be extracted from them}
\label{fig:example}
\end{center}
\end{figure*}

\section*{Background}

In principle, image mining involves the same processes as classical literature mining \cite{debruijn2002ijmi}: document categorization, named entity tagging, fact extraction, and collection-wide analysis. However, there are some subtle differences. Document categorization corresponds to image categorization, which is different in the sense that it has to deal with features based on the two-dimensional space of pixels, but otherwise the same principles of automatic categorization apply. Named entity tagging is different in two ways: pinpointing the mention of an entity is more difficult with images (a large number of pixels versus a couple of characters), and OCR errors have to be considered. Fact extraction in classical literature mining involves the analysis of the syntactic structure of the sentences. In images, in contrast, there are rarely complete sentences, but the semantics is rather encoded by graphical means. Thus, instead of parsing sentences, one has to analyze graphical elements and their relation to each other. The last process, collection-wide analysis, is a higher-level problem, and therefore no fundamental differences can be expected. Thus, image mining builds upon the same general stages as classical text mining, but with some subtle yet important differences.

Image mining on biomedical publications is not a new idea. It has been applied for the extraction of subcellular location information \cite{murphy2004ksce}, the detection of panels of fluorescence microscopy images \cite{qian2008bioinform}, the extraction of pathway information from diagrams \cite{kozhenkov2012bioinform}, and the detection of axis diagrams \cite{kuhn2012amia}. Also, there is a large amount of existing work on how to process gel images \cite{lemkin1997electrophoresis,luhn2003proteomics,cutler2003proteomics,rogers2003proteomics,zerr2005nar} and databases have been proposed to store the results of gel analyses \cite{schlamp2008gene}. These techniques, however, take as input plain gel images, which are not readily accessible from biomedical papers, because they make up just parts of the figures. Furthermore, these tools are designed for researchers who want to analyze their gel images and not to read gel diagrams that have already been analyzed and annotated by a researcher. Therefore, these approaches do not tackle the problem of recognizing and analyzing the labels of gel images. Some attempts to classify biomedical images include gel figures \cite{rodriguez2009bioinform}, which is, however, just the first step in locating them and analyzing their labels and their structure. To our knowledge, nobody has yet tried to perform image mining on gel diagrams.

\section*{Approach and Methods}

Figure \ref{fig:procedure} shows the procedure of our approach to image mining from gel diagrams. It consists of seven steps: figure extraction, segmentation, text recognition, gel detection, gel panel detection, named entity recognition and relation extraction.\footnote{Due to the fact that many figures consist of multiple panels of different types, we go straight to gel segment detection without first classifying entire images. Most gel panels share their figure with other panels, which makes automated classification difficult at the image level.}
\begin{figure*}[t]
\begin{center}
\includegraphics[width=0.75\textwidth]{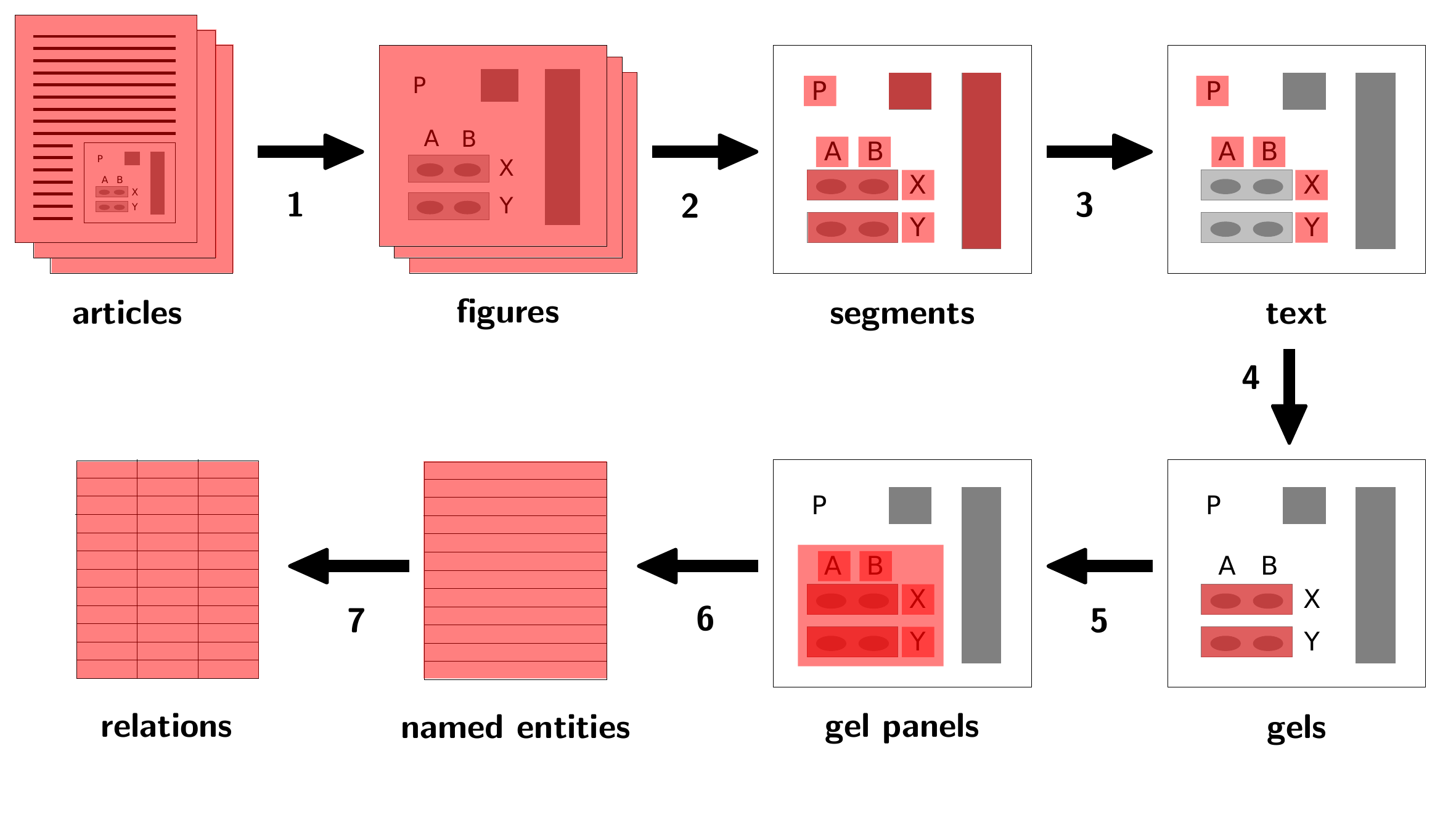}
\caption{The procedure of our approach: (1) figure extraction, (2) segmentation, (3) text recognition, (4) gel detection, (5) gel panel detection, (6) named entity recognition, and (7) relation extraction.}
\label{fig:procedure}
\end{center}
\end{figure*}

Using structured article representations, the first step is trivial. For steps two and three, we rely on existing work. The main focus of this paper lies on steps four and five: the detection of gels and gel panels. In the discussion section, we present some preliminary results on step six of recognizing named entities, and sketch how step seven could be implemented, for which we cannot provide a concrete solution at this point.

To practically evaluate our approach, we ran our pipeline on the entire open access subset of PubMed Central (though not all figures made it through the whole pipeline due to technical difficulties).

\subsection*{Figure Extraction}

A large portion of the articles of the open access subset of the PubMed Central database are available as structured XML files with additional image files for the figures. We only use these articles so far, which makes the figure extraction task very easy. It would be more difficult, though definitely feasible, to extract the figures from PDF files or even bitmaps of scanned articles (see \cite{ramakrishnan2012scbm} and \url{http://pdfjailbreak.com} for approaches on extracting the structure of articles in PDF format).

\subsection*{Segmentation and Text Recognition}

For the next two steps --- segment detection and subsequent text recognition ---, we rely on our previous work \cite{xu2010jbi,xu2011bsec}. This method includes the detection of layout elements, edge detection, and text recognition with a novel pivoting approach. For optical character recognition (OCR), the Microsoft Document Imaging package is used, which is available as part of Microsoft Office 2003. Overall, this approach has been shown to perform better than other existing approaches for the images found in biomedical publications \cite{xu2010jbi}. We do not go into the details here, as this paper focuses on the subsequent steps.

Due to some limitations of the segmentation algorithm when it comes to rectangles with low internal contrast (like gels), we applied a complementary very simple rectangle detection algorithm.

\subsection*{Gel Segment Detection}

Based on the results of the above-mentioned steps, we try to identify gel segments. Such gel segments typically have rectangular shapes with darker spots on a light gray background, or --- less commonly --- white spots on a dark background. We decided to use machine learning techniques to generate classifiers to detect such gel segments. To do so, we defined 39 numerical features for image segments: the coordinates of the relative position (within the image), the relative and absolute width and height, 16 grayscale histogram features, three color features (for red, green and blue), 13 texture features (coarseness, presence of ripples, etc.) based on \cite{haralick1973tsmc}, and the number of recognized characters.

To train the classifiers, we took a random sample of 500 figures, for which we manually annotated the gel segments. In the same way, we obtained a second sample of another 500 figures for testing the classifiers.\footnote{We double-checked these manual annotations to check their quality, which revealed only four misclassified segments in total for the training and test samples (0.016\% of all segments).}
We used the Weka toolkit and opted for random forest classifiers based on 75 random trees. Using different thresholds to adjust the trade-off between precision and recall, we generated a classifier with good precision and another one with good recall. Both of them are used in the next step.
We tried other types of classifiers including naive Bayes, Bayesian networks \cite{cooper1992ml}, PART decision lists \cite{frank1998icml}, and convolutional networks \cite{lecun1995handbook}, but we achieved the best results with random forests.

\subsection*{Gel Panel Detection}

A gel panel typically consists of several gel segments and comes with labels describing the involved genes, proteins, and conditions. For our goal, it is not sufficient to just detect the figures that contain gel panels, but we also have to extract their positions within the figures and to access their labels. This is not a simple classification task, and therefore machine learning techniques do not apply that easily. For that reason, we used a detection procedure based on hand-coded rules.

In a first step, we group gel segments to find contiguous gel regions that form the center part of gel panels. To do so, we start with looking for segments that our high-precision classifier detects as gel segments. Then, we repeatedly look for adjacent gel segments, this time applying the high-recall classifier, and merge them. Two segments are considered neighbors if they are at most 50 pixels apart\footnote{We are using absolute distance values at this point. A more refined algorithm could apply some sort of relative measure. However, the resolution of the images does not vary that much, which is why absolute values worked out well so far.} and do not have any text segment between them. Thus, segments which could be gel segments according to the high-recall classifier make it into a gel panel only if there is at least one high-precision segment in their group. The goal is to detect panels with high precision, but also to detect the complete panels and not just parts of them.
We focus here on precision because low recall can be leveraged by the large number of available gel images. Furthermore, as the open access part of PubMed Central only makes up a small subset of all biomedical publications, recall in a more general sense is anyway limited by the proportion of open access publications.

As a next step, we collect the labels in the form of text segments located around the detected gel regions. For a text segment to be attributed to a certain gel panel, its nearest edge must be at most 30 pixels away from the border of the gel region and its farthest edge must not be more than 150 pixels away. We end up with a representation of a gel panel consisting of two parts: a center region containing a number of gel segments and a set of labels in the form of text segments located around the center region.

To evaluate this algorithm, we collected yet another sample of 500 figures, in which 106 gel panels in 61 different figures were revealed by manual annotation.\footnote{Again, these manual annotations were double-checked to ensure their quality. Five errors were found and fixed in this process.} Based on this sample, we manually checked whether our algorithm is able to detect the presence and the (approximate) position of the gel panels.

\section*{Results}

The top part of Table \ref{tab:geldetection} shows the result of the gel detection classifier. We generated three different classifiers from the training data, one for each of the threshold values 0.15, 0.3 and 0.6. Lower threshold values lead to higher recall at the cost of precision, and vice versa. In the balanced case, we achieved an F-score of 75\%. To get classifiers with precision or recall over 90\%, F-score goes down significantly, but stays in a sensible range. These two classifiers (thresholds 0.15 and 0.6) are used in the next step. To interpret these values, one has to consider that gel segments are greatly outnumbered by non-gel segments. Concretely, only about 3\% are gel segments. More sophisticated accuracy measures for classifier performance, such as the area under the ROC curve \cite{bradley1997pr}, take this into account. For the presented classifiers, the area under the ROC curve is 98.0\% (on a scale from 50\% for a trivial, worthless classifier to 100\% for a perfect one).
\begin{table*}[t]
\begin{center}
\begin{tabular}{l|lr|rrrr}
 & Method & Threshold & Precision & Recall & F-score & ROC area \\
\hline
\multirow{7}{*}{Segments} & \multirow{3}{*}{Random forests} & 0.15 & 0.439 & 0.909 & 0.592 & \multirow{3}{*}{$\left.\begin{matrix}~\\~\\\end{matrix}\right\}$ 0.980} \\
 & & 0.30 & 0.765 & 0.739 & 0.752 \\
 & & 0.60 & 0.926 & 0.301 & 0.455 \\
\cline{2-7}
 & Naive Bayes & & 0.172 & 0.739 & 0.279 & 0.883 \\
 & Bayesian network & & 0.394 & 0.531 & 0.452 & 0.914 \\
 & PART decision list & & 0.631 & 0.496 & 0.555 & 0.777 \\
 & Convolutional networks & & 0.142 & 0.949 & 0.248 & \\
\hline
Panels & Hand-coded rules & & 0.951 & 0.368 & 0.530 & \\
\end{tabular}
\caption{The results of the gel segment detection classifiers (top) and the gel panel detection algorithm (bottom)}
\label{tab:geldetection}
\end{center}
\end{table*}

The results of the gel panel detection algorithm are shown in the bottom part of Table \ref{tab:geldetection}. The precision is 95\% at a recall of 37\%, leading to an F-score of 53\%. The comparatively low recall is mainly due to the general problem of pipeline-based approaches that the various errors from the earlier steps accumulate and are hard to correct at a later stage in the pipeline.

Table \ref{tab:pipeline} shows the results of running the pipeline on PubMed Central. We started with about 410\,000 articles, the entire open access subset of PubMed Central at the time we downloaded them (February 2012). We successfully parsed the XML files of 94\% of these articles (for the remaining articles, the XML file was missing or not well-formed, or other unexpected errors occurred). The successful articles contained around 1\,100\,000 figures, for some of which our segment detection step encountered image formatting errors or other internal errors, or was just not able to detect any segments. We ended up with more than 880\,000 figures, in which we detected about 86\,000 gel panels, i.e. roughly ten out of 100 figures. For each of them, we found on average 3.6 labels with recognized text. After tokenization, we identified about 76\,000 gene names in these gel labels, which corresponds to 6.8\% of the tokens. Considering all text segments (including but not restricted to gel labels), only 3.3\% of the tokens are detected as gene names.\footnote{The low numbers are partially due to the fact that a considerable part of the tokens are ``junk tokens'' produced by the OCR step when trying to recognize characters in segments that do not contain text.}
\begin{table*}[t]
\begin{center}
\begin{tabular}{l|r}
Total articles & 410\,950 \\
Processed articles & 386\,428 \\
\hline
Total figures from processed articles & 1\,110\,643 \\
Processed figures & 884\,152 \\
\hline
Detected gel panels & 85\,942 \\
Detected gel panels per figure & 0.097 \\
Detected gel labels & 309\,340 \\
Detected gel labels per panel & 3.599 \\
\hline
Detected gene tokens & 1\,854\,609 \\
Detected gene tokens in gel labels & 75\,610 \\
Gene token ratio & 0.033 \\
Gene token ratio in gel labels & 0.068 \\
\end{tabular}
\caption{The results of running the pipeline on the open access subset of PubMed Central}
\label{tab:pipeline}
\end{center}
\end{table*}

\section*{Discussion}

The presented results show that we are able to detect gel segments with high accuracy, which allows us to subsequently detect whole gel panels at a high precision. The recall of the panel detection step is relatively low, but with about 37\% still in a reasonable range. As mentioned above, we can leverage the high number of available figures, which makes precision more important than recall. Running our pipeline on the whole set of open access articles from PubMed Central, we were able to retrieve 85\,942 potential gel panels (around 95\% of which we can expect to be correctly detected).

The next step would be to recognize the named entities mentioned in the gel labels. To this aim, we did a preliminary study to investigate whether we are able to extract the names of genes and proteins from gel diagrams. To do so, we tokenized the label texts and looked for entries in the Entrez Gene database to match the tokens. This look-up was done in a case-sensitive way, because many names in gel labels are acronyms, where the specific capitalization pattern can be critical to identify the respective entity. We excluded tokens that have less than three characters, are numbers (Arabic or Latin), or correspond to common short words (retrieved from a list of the 100 most frequent words in biomedical articles). In addition, we extended this exclusion list with 22 general words that are frequently used in the context of gel diagrams, some of which coincide with gene names according to Entrez.\footnote{These words are: \emph{min}, \emph{hrs}, \emph{line}, \emph{type}, \emph{protein}, \emph{DNA}, \emph{RNA}, \emph{mRNA}, \emph{membrane}, \emph{gel}, \emph{fold}, \emph{fragment}, \emph{antigen}, \emph{enzyme}, \emph{kinase}, \emph{cleavage}, \emph{factor}, \emph{blot}, \emph{pro}, \emph{pre}, \emph{peptide}, and \emph{cell}.}
Since gel electrophoresis is a method to analyze genes and proteins, we would expect to find more such mentions in gel labels than in other text segments of a figure. By measuring this, we get an idea of whether the approach works out or not. In addition, we manually checked the gene and protein names extracted from gel labels after running our pipeline on 2\,000 random figures. In 124 of these figures, at least one gel panel was detected.
Table \ref{tab:geneprecision} shows the results of this preliminary evaluation. Almost two-thirds of the detected gene/protein tokens (65.3\%) were correctly identified. 9\% thereof were correct but could be more specific, e.g. when only ``actin'' was recognized for ``$\beta$-actin'' (which is not incorrect but of course much harder to map to a meaningful identifier). The incorrect cases (34.6\%) can be split into two classes of roughly the same size: some recognized tokens were actually not mentioned in the figure but emerged from OCR errors; other tokens were correctly recognized but incorrectly classified as gene or protein references.
\begin{table*}[t]
\begin{center}
\begin{tabular}{l|rr}
& absolute & relative \\
\hline
\textbf{Total} & \textbf{156} & \textbf{100.0\%} \\
\hline
\textbf{Incorrect} & \textbf{54} & \textbf{34.6\%} \\
-- Not mentioned (OCR errors) & 28 & 17.9\% \\
-- Not references to genes or proteins & 26 & 16.7\% \\
\hline
\textbf{Correct} & \textbf{102} & \textbf{65.3\%} \\
-- Partially correct (could be more specific) & 14 & 9.0\% \\
-- Fully correct & 88 & 56.4\% \\
\end{tabular}
\caption{Numbers of recognized gene/protein tokens in 2\,000 random figures}
\label{tab:geneprecision}
\end{center}
\end{table*}
Although there is certainly much room for improvement, this simple gene detection step seems to perform reasonably well.

For the last step, relation extraction, we cannot present any concrete results at this point. After recognizing the named entities, we would have to disambiguate them, identify their semantic roles (condition, measurement or something else), align the gel images with the labels, and ultimately quantify the degree of expression. To improve the quality of the results, combinations with classical text mining techniques should be considered. This is all future work. We expect to be able to profit to a large extent from existing work to disambiguate protein and gene names \cite{tanabe2002bioinform,lu2011bmcbioinf} and to detect and analyze gel spots (see the existing work mentioned above).

It seems reasonable to assume that these results can be combined with existing techniques of term disambiguation and gel spot detection at a satisfactory level of accuracy. We plan to investigate this in future work.

As mentioned above, we have started to investigate how the gel segment detection step could be improved by the use of the image recognition technique of convolutional networks (ConvNet) \cite{lecun1995handbook}. We started with a simplified approach to the one presented in \cite{barbano2013bsec}. In this approach, images are tiled into small quadratic pieces. We used a single network (and not several parallel networks), based on $48 \times 48$ input tile images with three layers of convolutions. The first layer takes eight $5 \times 5$ convolutions and is followed by a $2 \times 2$ sub-sampling. The second layer takes twenty four $5 \times 5$ convolutions and is followed by a $3 \times 3$ sub-sampling. The last layer takes seventy two $6 \times 6$ convolutions, which leads to a fully connected layer.
We trained our ConvNet on the 500 images of the training set, where we manually annotated the tiles as \emph{gel} and \emph{non-gel}.
With the use of EBLearn \cite{sermanet2009ictai}, this trained ConvNet classified the tiles of the 500 images of our testing set.
The classified tiles can then be reconstructed into a mask image, as shown in Figure \ref{fig:convnet}. A manual check of the clusters of recognized gel tiles led to the results shown in Table \ref{tab:geldetection}. Recall is very good (95\%) but precision is very poor (14\%), leading to an F-score of 25\%. This is much worse than the results we got with our random forest approach, which is why ConvNet is currently not part of our pipeline. We hope, however, that we can further optimize this ConvNet approach and combine it with random forests to exploit their (hopefully) complementary benefits. Using ConvNet to classify complete images as \emph{gel-image} or \emph{non-gel-image} and adjusting the classification to account for unbalanced classes, we were able to obtain an F-score of 74\%, which makes us confident that a combination of the two approaches could lead to a significant improvement of our gel segment detection step.
As an alternative approach, we will try to run ConvNet on down-scaled entire panels rather than small tiles, as described in \cite{krizhevsky2012nips}. Furthermore, we will experiment with parallel networks instead of single ones to improve accuracy.

\begin{figure*}[tb]
\begin{center}
\setlength{\fboxsep}{0pt}
\framebox{\includegraphics[height=70mm]{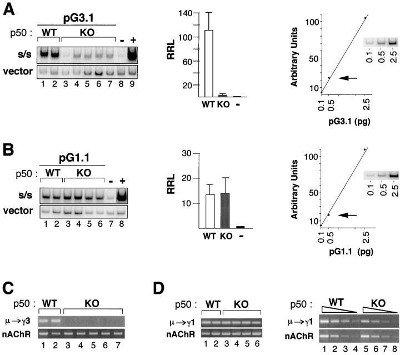}}
\hfill
\framebox{\includegraphics[height=70mm]{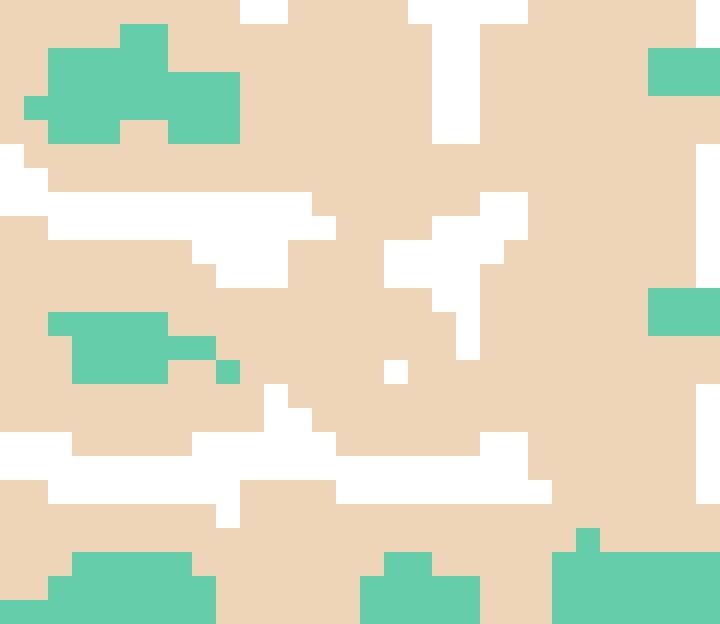}}
\caption{Original and mask image after ConvNet classification for an exemplary image from PMID 14993249. Green means \emph{gel}; brown means \emph{other}; and white means \emph{not enough gradient information}.}
\label{fig:convnet}
\end{center}
\end{figure*}

The results obtained from our gel recognition pipeline indicate that it is feasible to extract relations from gel images, but it is clear that this procedure is far from perfect. The automatic analysis of bitmap images seems to be the only efficient way to extract such relations from existing publications, but other publishing techniques should be considered for the future. The use of vector graphics instead of bitmaps would already greatly improve any subsequent attempts of automatic analysis. A further improvement would be to establish accepted standards for different types of biomedical diagrams in the spirit of the Unified Modeling Language, a graphical language widely applied in software engineering since the 1990s. Ideally, the resulting images could directly include semantic relations in a formal notation, which would make relation mining a trivial procedure. If authors are supported by good tools to draw diagrams like gel images, this approach could turn out to be feasible even in the near future.

Concretely, we would like to take the opportunity to postulate the following actions, which we think should be carried out to make the content of images in biomedical articles more accessible:
\begin{itemize}
\item \textbf{Stop pressing diagrams into bitmaps!} Unless the image only consists of one single photograph, screenshot, or another kind of picture that only has bitmap representation, vector graphics should be used for article figures.
\item \textbf{Let data and metadata travel from the tools that generate diagrams to the final articles!} Whenever the specific tool that is used to generate the diagram ``knows'' that a certain graphical element refers to an organism, a gene, an interaction, a point in time, or another kind of entity, then this information should be stored in the image file, passed on, and finally published with the article.
\item \textbf{Use RDF vocabularies to embed semantic annotations in diagrams!} Tools for creating scientific diagrams should use RDF notation and stick to existing standardized schemas (or define new ones if required) to annotate the diagram files they create.
\item \textbf{Define standards for scientific diagrams!} In the spirit of the Unified Modeling Language, the biomedical community should come up with standards that define the appearance and meaning of different types of diagrams.
\end{itemize}
Obviously, different groups of people need to be involved in these actions, namely article authors, journal editors, and tool developers. It is relatively inexpensive to follow these postulates (though it might require some time), which in turn would greatly improve data sharing, image mining, and scientific communication in general. Standardized diagrams could be the long sought solution to the problem of how to let authors publish computer-processable formal representations for (part of) their results.
This can build upon the efforts of establishing an open annotation model \cite{ciccarese2011jbiomedsem,sanderson2013openannotation}.

\section*{Conclusions}

Successful image mining from gel diagrams in biomedical publications would unlock a large amount of valuable data. Our results show that gel panels and their labels can be detected with high accuracy, applying machine learning techniques and hand-coded rules. We also showed that genes and proteins can be detected in the gel labels with satisfactory precision.

Based on these results, we believe that this kind of image mining is a promising and viable approach to provide more powerful query interfaces for researchers, to gather relations such as protein-protein interactions, and to generally complement existing text mining approaches. At the same time, we believe that an effort towards standardization of scientific diagrams such as gel images would greatly improve the efficiency and precision of image mining at relatively low additional costs at the time of publication.

\ifthenelse{\boolean{publ}}{\small}{}

\section*{Competing Interests}

The authors declare that they have no competing interests.

\section*{Authors' Contributions}

TK was the main author and main contributor of the presented work. He was responsible for designing and implementing the pipeline, gathering the data, performing the evaluation, and analyzing the results. MLN applied, trained, and evaluated the ConvNet classifier, and contributed to the annotation of the test sets. TL built and analyzed the corpus for the preparatory evaluation. MK contributed to the conception and the design of the approach and to the analysis of the results. All authors have been involved in drafting or revising the manuscript, and all authors read and approved the final manuscript.

\section*{Acknowledgments}

This study has been supported by the National Library of Medicine grant 5R01LM009956.

\newpage
{
\ifthenelse{\boolean{publ}}{\footnotesize}{\small}
\bibliographystyle{bmc_article}
\bibliography{gels}
}

\ifthenelse{\boolean{publ}}{\end{multicols}}{}

\end{bmcformat}
\end{document}